\documentclass[12pt]{article}
\usepackage{amsfonts,amssymb,amsmath}
\usepackage{epic,eepic}
\usepackage{theorem,cite}
\theorembodyfont{\rmfamily}

\numberwithin{equation}{section}

          \def\cB{{\cal B}}          
          \def\cE{{\cal E}}          
          \def\cH{{\cal H}}          
\def\cJ{{\cal J}}                    
                    \def\cO{{\cal O}}

\def\fg{{\mathfrak g}}

\newcommand{\CC}{{\mathbb C}}

\newcommand{\qmbox}[1]{{\qquad\mbox{#1}\quad}}
\def\qmbox#1{\qquad\mbox{#1}\quad}

\begin{document}
\pagestyle{empty}

\null

\begin{center}

{\Large \textsf{New spin generalisation for \\
long range interaction models }}

\vspace{10mm}

{\large N.~Cramp{\'e}}

\vspace{5mm}

\emph{University of York, Department of Mathematics}

\emph{Heslington, York YO10 5DD, United Kingdom}

\emph{nc501@york.ac.uk}

\vfill
\end{center}
\vfill  \rightline{\textit{Dedicated to my PhD supervisor and friend
D. Arnaudon}} \vfill

\begin{abstract}
We study new interactions between degrees of freedom for Calogero,
Sutherland and confined Calogero spin models. These interactions are
encoded by the generators of the Lie algebra $so(N)$ or $sp(N)$. We
find the symmetry algebras of these new models: the half-loop
algebra based on $so(N)$ or $sp(N)$ for the Calogero models and the
Yangian of $so(N)$ or $sp(N)$ for the two types of other models.
Surprisingly, these symmetry occur only for a specific value of the
coupling constant.

\end{abstract}
\vfill \vfill MSC: 70H06, 81R12, 81R50 --- PACS: 02.20.Uw, 02.30.Ik,
03.65.Fd, 75.10.Pq \vspace{5mm} \vfill
\rightline{\today}

\baselineskip=16pt

\newpage

\pagestyle{plain} \setcounter{page}{1}

\section*{Introduction}

\quad~The Calogero and Sutherland models are one-dimensional many
body problems with long range interactions \cite{Cal,Sut}. The
introduction of a $gl(N)$ internal degree of freedom in these models
\cite{MP,Poly,HH} has proved to be fruitful in various physical and
mathematical investigations. This is well illustrated in the study
of their symmetries which turn out to be the half-loop algebra or
the Yangian associated to $gl(N)$ \cite{BGHP,BHW,Hik1,F01_1}. This
letter is devoted to the introduction of new interactions between
the internal degrees of freedom in these models and finding the
symmetry algebra of these new models. The interactions are defined
thanks to the fundamental representation of the generators of the
Lie algebras $so(N)$ or $sp(N)$. We will call these new models
$so(N)$ or $sp(N)$-spin models. At this point, to avoid ambiguity,
let us remark that these models are different from the so-called
$BC_N$ models \cite{OP}. Indeed, in the latter case, it is the
potential which is closely related to the root systems of the
algebra $BC_N$ and such models possess reflection algebra symmetry
\cite{caduk}.

The plan of the letter is as follows. In section \ref{sect:general},
we introduce the definitions and different notations used in the
letter. Then, in the three following sections which have the same
structure, we introduce the Hamiltonian for Calogero $so(N)$ or
$sp(N)$ spin model, Sutherland $so(N)$ or $sp(N)$ spin model and
confined Calogero $so(N)$ or $sp(N)$ spin model. The main results of
this letter consists in finding for each model the symmetry algebra.
We finish by an appendix where technical details for computations
are gathered.

\section{General setting\label{sect:general}}
\setcounter{equation}{0}

Let $x_1,\dots,x_L$ be the positions of $L$ particles on a
one-dimensional space. We associate to each particle an internal
degree of freedom or spin which will be considered as a vector
belonging to $\CC^N$. The spin operators $E^{ab}_i$ ($1\leqslant a,b
\leqslant N$ and $1\leqslant i \leqslant L$) are matrices with entry
1 in row $a$ and column $b$ and zero elsewhere which act on the spin
space of the $i^{th}$ particle. They provide a representation of
$\oplus_{1}^{L} gl(N)$ and they satisfy the following commutation
relations
\begin{equation}
[E^{ab}_i,E^{cd}_j]=\delta_{ij}\left(
\delta_{bc}\,E^{ad}_i-\delta_{ad}\,E^{cb}_i
 \right)\;.
\end{equation}
Let $\theta_0=\pm 1$. For each index $1\leqslant a \leqslant N$, we
introduce the following sign, for $N$ even,
  \begin{equation}
    \label{eq:deftheta}
        \theta_a = \begin{cases}
      +1 & \qmbox{for} 1\leqslant a\leqslant \frac{N}{2}\;,  \cr
      \theta_0 & \qmbox{for} \frac{N}{2}+1 \leqslant a\leqslant N\;,
    \end{cases}
  \end{equation}
and $\theta_a=+1$ if N is odd. We introduce also the following
conjugate index $\bar{a}$
  \begin{equation}
    \label{eq:defibar}
    \bar{a} =N+1-a \qmbox{for} 1\leqslant a\leqslant N\;.
  \end{equation}
In particular $\theta_{a} \theta_{\bar{a}} = \theta_{0}$.

These definitions allow us to deal simultaneously with the Lie
algebras $so(N)$ and $sp(N)$, subalgebras of $gl(N)$. Let us define,
for $1\leqslant a,b \leqslant N$,
\begin{equation}
\label{eq:defF} F^{ab}=E^{ab}-\theta_a\theta_b E^{\bar{b}\bar{a}}\;.
\end{equation}
The algebra $\fg^+(N)$ (resp. $\fg^-(N)$) spanned by these
generators is isomorphic to $so(N)$ (resp. $sp(N)$) when
$\theta_0=+1$ (resp. $\theta_0=-1$). Of course, the case
$\theta_0=-1$ occurs only when $N$ is even. These generators satisfy
the symmetry relation
\begin{equation}
\label{eq:symF} F^{ab}=-\theta_a\theta_bF^{\bar{b}\bar{a}}\;.
\end{equation}
To correctly define the structure constants and a non-degenerate
metric tensor of $\fg^\pm(N)$, we need to restrict this set
$\{F^{ab}\}$ of generators to a basis of $\fg^\pm(N)$. Let us define
the subsets of indices $\cE^+=\{(a,b)|\bar a>b\}$ and
$\cE^-=\{(a,b)|\bar a\geqslant b\}$. The sets
$\cB^\pm=\{F^{ab}|(a,b)\in \cE^\pm\}$ form bases of Lie algebras
$\fg^\pm(N)$. Then, the commutation relations of $\fg^\pm(N)$ can be
written as follows, for $(a,b),(c,d)\in\cE^\pm$,
\begin{eqnarray}
\label{eq:comF} [F^{ab}_i,F^{cd}_j]=\delta_{ij}~\sum_{(e,f)\in
\cE^\pm}{f^{ab,cd}}_{ef}~F^{ef}_i\;,
\end{eqnarray}
where the structure constants read
\begin{eqnarray}
\label{eq:const1}
{f^{ab,cd}}_{ef}&=&\big[\delta^{bc}(\delta^a_e\delta^d_f-
\theta_a\theta_d\delta^a_{\bar f}\delta^d_{\bar e})
-\delta^{ad}(\delta^b_f\delta^c_e-\theta_b\theta_c\delta^b_{\bar
e}\delta^c_{\bar f}) \nonumber\\
&& -\delta^{a\bar c}(\theta_a\theta_b\delta^b_{\bar
e}\delta^d_f-\theta_c\theta_d\delta^b_{f}\delta^d_{\bar
e})+\delta^{b \bar d}(\theta_a\theta_b\delta^a_{\bar
f}\delta^c_e-\theta_c\theta_d\delta^a_{e}\delta^c_{\bar f})
\big]H(\bar e,f)\;.\hspace{1cm}
\end{eqnarray}
The function $H(i,j)$ is defined, for $(i,j)\in \cE^\pm$, as follows
\begin{eqnarray}
H(i,j)= \begin{cases}1\qmbox{if} i>j\;,\\
\frac{1}{2} \qmbox{if}i=j\;. \end{cases}
\end{eqnarray}
The factor $1/2$ in the function $H$ is relevant only in the case
where we consider $\fg^-(N)$ and is due to the particular choice of
the normalisation of the generators. We choose the non-degenerate
metric tensor as follows, for $(a,b),(c,d)\in \cE^\pm$,
\begin{eqnarray}
\label{eq:metric} g^{ab,cd}=\frac{1}{2} \mbox{Tr}(F^{ab}F^{cd})\;.
\end{eqnarray}
This metric will allow us to raise or lower the indices of the
structure constants.

\section{Calogero model}

In this section, we will obtain the symmetry algebra of the Calogero
$\fg^\pm(N)$-spin model which we defined through the following
Hamiltonian
\begin{equation}
\label{eq:hamcal} \cH_C=-\sum_{j=1}^L \frac{\partial^2}{\partial
x_j^2} + \sum_{j\neq k} \frac{\lambda^2-\lambda P_{jk}+ \lambda
Q_{jk}}{(x_j-x_k)^2}\;.
\end{equation}
The matrix $P_{jk}$ permutes the spins of the j$^{th}$ and k$^{th}$
particles and can be written in terms of the spin operators as
  \begin{equation}
    \label{eq:Pdef}
    P_{jk} = \sum_{a,b=1}^{N} E^{ab}_j E^{ba}_k\;.
  \end{equation}
The operator $Q_{jk}$ is defined by
  \begin{equation}
    \label{eq:Qdef}
    Q_{jk} = \sum_{a,b=1}^{N}  \theta_{a} \theta_{b}
    E^{ab}_j E^{\bar{a}\bar{b}}_k \;.
  \end{equation}
They satisfy, in particular, the useful properties $P_{jk}=P_{kj}$
and $Q_{jk}=Q_{kj}$. These two operators are the crucial elements to
construct the R-matrix associated to the Yangian of $so(N)$ or
$sp(N)$ \cite{ABZABZ78,KS,Isa,soya}.

The introduction in the Hamiltonian of the operator $Q_{jk}$
modifies the interaction between the degrees of freedom of the
particles as compared with the $gl(N)$-spin model. Note that we can
write the new interactions in terms of the generators of
$\fg^\pm(N)$ as follows
\begin{equation}
P_{jk}-Q_{jk}=\frac{1}{2}\sum_{a,b=1}^N F_j^{ab}F_k^{ba}\;.
\end{equation}
We have used in the previous formula the conventional notation
$(F_jF_k)^{ab}=\sum_{c=1}^NF_j^{ac}F_k^{cb}$.

It is well-known that the symmetry algebra of $A_N$ Calogero
$gl(N)$-spin model is the half-loop algebra of $gl(N)$
\cite{BGHP,BHW,Hik1,F01_1}. We shall show how this symmetry algebra
is modified for Calogero $\fg^\pm(N)$-spin model. We introduce the
following operators, for $(a,b)\in \cE^\pm$,
\begin{eqnarray}
\label{eq:gen_cal0}
 J_0^{ab}&=&\sum_{j=1}^L F_j^{ab}\;,\\
 \label{eq:gen_cal1}
 J_1^{ab}&=&\sum_{j=1}^L F_j^{ab}\frac{\partial}{\partial x_j}
 -\lambda \sum_{j\neq k} (F_jF_k)^{ab}\frac{1}{x_j-x_k}\;.
\end{eqnarray}
After a straightforward computation, we can show that these
operators satisfy the following relations
\begin{eqnarray}
\label{eq:comJ0} &&[J^{ab}_0,J^{cd}_0]=
{f^{ab,cd}}_{ef}~J^{ef}_0\;,\\
\label{eq:comJ1} &&[J^{ab}_0,J^{cd}_1]=
{f^{ab,cd}}_{ef}~J^{ef}_1\;, \\
\label{eq:serrecal} &&
\Big[J_1^{ab},\big[J_0^{cd},J_1^{ef}\big]\Big]
+\Big[J_1^{ef},\big[J_0^{ab},J_1^{cd}\big]\Big]
+\Big[J_1^{cd},\big[J_0^{ef},J_1^{ab}\big]\Big]=0\;,
\end{eqnarray}
for the following particular value of the coupling constant
\begin{equation}
\label{eq:lambda}
\lambda=\frac{2}{N-4\theta_0}\;.
\end{equation}
In relations (\ref{eq:comJ0}) and (\ref{eq:comJ1}), we have used the
Einstein's notation for the repeated pair of indices $(e,f)$ but the
sums are only for $(e,f)\in \cE^\pm$ (for example, we write
explicitly this sum in (\ref{eq:comF})). From now on, we use this
convention for the repeated indices.

For the particular choice $N=4\theta_0$, the denominator in
(\ref{eq:lambda}) vanishes and therefore $J^{ab}_0$, $J^{ab}_1$ are
not well-defined. However, this corresponds to the case where we
consider the algebra $so(4)$ which is a non-simple Lie algebra.

The higher level generators, $J_2^{ab}, J_3^{ab},\dots$, are defined
recursively from $J_0^{ab}$ and $J_1^{ab}$. Relation
(\ref{eq:serrecal}), called Serre relation, guarantees that these
generators are well-defined. We have the commutation relations
\begin{equation}
\label{eq:comJn}
[J^{ab}_n,J^{cd}_m]={f^{ab,cd}}_{ef}~J^{ef}_{n+m}\;.
\end{equation}
Relations (\ref{eq:comJ0})-(\ref{eq:serrecal}) define the half-loop
algebra (also called Gaudin algebra) associated to the Lie algebra
$\fg^\pm(N)$.

To finish the proof of the symmetry, we show that $J_0^{ab}$ and
$J_1^{ab}$ are conserved operators i.e.
\begin{equation}
\label{eq:hamJ} [\cH_C,J_0^{ab}]=0 \qmbox{and} [\cH_C,J_1^{ab}]=0\;.
\end{equation}
The particular value (\ref{eq:lambda}) of $\lambda$ is necessary and
sufficient to prove the second relation in (\ref{eq:hamJ}) whereas
the first one holds for any value of $\lambda$. We have used, in
particular, the two following properties
\begin{equation}
P_{jk}F_j^{ab}=F_k^{ab}P_{jk}\qmbox{and}
Q_{jk}F_j^{ab}=-Q_{jk}F_k^{ab}\;.
\end{equation}

Therefore, we have therefore shown that the symmetry algebra of the
model described by the Hamiltonian (\ref{eq:hamcal}) is the
half-loop algebra associated to $\fg^\pm(N)$. To obtain the
symmetry, it was necessary to constrain the coupling constant. This
feature is new in comparison with the Calogero $gl(N)$-spin model
where the coupling constant remains arbitrary free.

\section{Sutherland model}

In this section, we introduce a new Sutherland spin model, called
Sutherland $\fg^\pm(N)$-spin model, whith Hamiltonian given by
\begin{equation}
\label{eq:hamsut} \cH_S=-\sum_{j=1}^L
\left(x_j\frac{\partial}{\partial x_j}\right)^2 + \sum_{j\neq k}
(\lambda^2-\lambda P_{jk}+ \lambda Q_{jk})\frac{x_jx_k}{(x_j-x_k)^2}
\end{equation}
and exhibit its symmetry algebra.

It is well-known that the symmetry algebra of Sutherland
$gl(N)$-spin model is the Yangian of $gl(N)$ and we show that for
this model it is the Yangian of $\fg^\pm(N)$. The end of this
section consists in proving this statement. For convenience, let us
define the symmetriser of any three elements $K^A,K^B,K^C$ by
\begin{equation}
\label{eq:sym} \{K^A,K^B,K^C\}=\frac{1}{24}\sum_{\sigma\in S_3}
K^{\sigma(A)}K^{\sigma(B)}K^{\sigma(C)}\;,
\end{equation}
where $S_3$ is the group of the permutations of order $6$. The
Yangian of $\fg^\pm(N)$ is the associative algebra generated by
$\{K^{ab}_0, K^{ab}_1|(a,b)\in\cE^\pm\}$ constrained by the
following commutation relations \cite{Dri85}, for $(a,b),(c,d)\in
\cE^\pm,$
\begin{eqnarray}
\label{eq:comK0} [K^{ab}_0,K^{cd}_0]&=&{f^{ab,cd}}_{ef}~K^{ef}_0\;,\\
\label{eq:comK1} [K^{ab}_0,K^{cd}_1]&=&{f^{ab,cd}}_{ef}~K^{ef}_1\;,
\end{eqnarray}
and by the Serre relations, for $(a,b),(c,d),(e,f)\in \cE^\pm,$
\begin{eqnarray}
&&\Big[K_1^{ab},\big[K_0^{cd},K_1^{ef}\big]\Big]
+\Big[K_1^{ef},\big[K_0^{ab},K_1^{cd}\big]\Big]
+\Big[K_1^{cd},\big[K_0^{ef},K_1^{ab}\big]\Big]\nonumber\\
\label{eq:serresut}&&\hspace{2cm}=\lambda^2
{f^{ab}}_{\alpha\beta,ij}{f^{cd}}_{\gamma\delta,kl}
{f^{ef}}_{\epsilon\phi,mn}{f^{ij,kl,mn}}
\{K_0^{\alpha\beta},K_0^{\gamma\delta},K_0^{\epsilon\phi}\}\;.
\end{eqnarray}
We recall that we use the Einstein's notation for repeated indices
but that the skipped sums are running over $\cE^\pm$. The Lie
algebra indices are lowered or raised by the invariant
non-degenerate metric tensor defined by relation (\ref{eq:metric}).

The following operators, for $(a,b)\in\cE^\pm$,
\begin{eqnarray}
\label{eq:gen_sut0}
 K_0^{ab}&=&\sum_{j=1}^L F_j^{ab}\;,\\
 \label{eq:gen_sut1}
 K_1^{ab}&=&\sum_{j=1}^L F_j^{ab}x_j\frac{\partial}{\partial x_j}
 -\lambda \sum_{j\neq k} (F_jF_k)^{ab}\frac{x_j+x_k}{x_j-x_k}\;,
\end{eqnarray}
give a representation of the Yangian of $\fg^\pm(N)$ provided that
the coupling constant $\lambda$ (which is also the deformation
parameter of the Yangian) takes the particular value
(\ref{eq:lambda}). The first two relations (\ref{eq:comK0}) and
(\ref{eq:comK1}) are easily proven by direct computation. We give
some details for the computation of the Serre relation
(\ref{eq:serresut}) in the appendix.

By direct computation, we can also prove that
\begin{equation}
\label{eq:hamK} [\cH_S,K_0^{ab}]=0 \qmbox{and} [\cH_S,K_1^{ab}]=0\;.
\end{equation}
The first relation in (\ref{eq:hamK}) is true for any $\lambda$
whereas the second one holds only and only if $\lambda$ is equal to
the particular value (\ref{eq:lambda}). Therefore, we have proved
that the symmetry algebra of the model described by the Hamiltonian
(\ref{eq:hamcal}) is the Yangian of $\fg^\pm(N)$ (for the
deformation parameter equal to
$\frac{2}{N-4\theta_0}$).\\

\section{Confined Calogero model}

This section is devoted to studying the symmetry algebra of the
confined Calogero $\fg^\pm(N)$-spin model which is described by the
following Hamiltonian
\begin{equation}
\label{eq:hamcalconf} \cH_{CC}=\cH_C+\omega^2\sum_{j=1}^L x_j^2\;.
\end{equation}
The operator $\cH_C$ is the Hamiltonian of the Calogero model
introduced in (\ref{eq:hamcal}). Let us remark that the introduction
of this harmonic potential breaks translation invariance.

We shall prove that the linear combinations introduced in \cite{BHW}
to obtain the symmetry algebra of the confined Calogero $gl(N)$-spin
chain model are also relevant in our case to obtain the symmetry
algebra. Let us define a new set of operators, for $(a,b)\in
\cE^{\pm}$
\begin{eqnarray}
\cJ^{ab}_0&=&J^{ab}_0\;,\\
\cJ^{ab}_1&=&J^{ab}_2-\omega^2 \cO^{ab}_2\;,
\end{eqnarray}
where we have introduced the new operators $\cO^{ab}_n=\sum_{j=1}^L
F_j^{ab}x_j^n$. We can easily show that the set of operators
$\{\cO^{ab}_n\}$ satisfy the relations of the half-loop algebra
(\ref{eq:comJn}). By computing $[J_1^{ab},J_1^{cd}]$, we find the
following explicit form for $J^{ab}_2$,
\begin{eqnarray}
J^{ab}_2&=&\sum_{j=1}^L F_j^{ab}\frac{\partial^2}{\partial x_j^2}
-\lambda\sum_{j\neq
k}\frac{(F_jF_k)^{ab}}{x_j-x_k}\left(\frac{\partial}{\partial
x_j}-\frac{\partial}{\partial x_k}\right) +\lambda\sum_{j\neq
k}\frac{(E_jE_k)^{ab}-\theta_a\theta_b(E_jE_k)^{\bar b \bar
a}-\lambda F_j^{ab}}{(x_j-x_k)^2}\nonumber\\
\label{eq:J2} &&-\lambda^2 \sum_{j\neq k \neq
\ell}\frac{(F_kF_jF_\ell)^{ab}}{(x_j-x_k)(x_j-x_\ell)}\;,
\end{eqnarray}
where we have used this following contraction $\displaystyle
(F_kF_jF_\ell)^{ab}=\sum_{\alpha,\beta=1}^N F_k^{a\alpha}
F_j^{\alpha\beta}F_\ell^{\beta b}$. We can prove that the operators
$\cJ^{ab}_0$ and $\cJ^{ab}_1$ provide a representation of the
Yangian of $\fg^\pm(N)$
\begin{eqnarray}
\label{eq:comcJ0} &&[\cJ^{ab}_0,\cJ^{cd}_0]={f^{ab,cd}}_{ef}~\cJ^{ef}_0\;,\\
\label{eq:comcJ1}&&
[\cJ^{ab}_0,\cJ^{cd}_1]={f^{ab,cd}}_{ef}~\cJ^{ef}_1\;,\\
&&\Big[\cJ_1^{ab},\big[\cJ_0^{cd},\cJ_1^{ef}\big]\Big]
+\Big[\cJ_1^{ef},\big[\cJ_0^{ab},\cJ_1^{cd}\big]\Big]
+\Big[\cJ_1^{cd},\big[\cJ_0^{ef},\cJ_1^{ab}\big]\Big]\nonumber\\
\label{eq:serreconfcal}&&\hspace{2cm}=4\lambda^2\omega^2
{f^{ab}}_{\alpha\beta,ij}{f^{cd}}_{\gamma\delta,kl}
{f^{ef}}_{\epsilon\phi,mn}{f^{ij,kl,mn}}
\{\cJ_0^{\alpha\beta},\cJ_0^{\gamma\delta},\cJ_0^{\epsilon\phi}\}\;.
\end{eqnarray}
Once again, these relations are satisfied if and only if the
parameter $\lambda$ takes the particular value (\ref{eq:lambda})
whereas the parameter $\omega$ remains free. Obviously, in the limit
$\omega$ tends to $0$, we recover the half-loop algebra for the
corresponding Calogero model. The commutation relations
(\ref{eq:comcJ0}) and (\ref{eq:comcJ1}) are computed directly. Some
details for the computation of the Serre relations are presented in
the appendix.

The proof of the symmetry is provided by the two following relations
\begin{equation}
\label{eq:hamcj} [\cH_{CC},\cJ_0^{ab}]=0 \qmbox{and}
[\cH_{CC},\cJ_1^{ab}]=0\;.
\end{equation}
Let us remark that the first relation in (\ref{eq:hamcj}) holds for
any $\lambda$ whereas the second one requires that $\lambda$ takes
the particular value (\ref{eq:lambda}). Therefore we have proved
that the symmetry of the model described by the Hamiltonian
(\ref{eq:hamcalconf}) is the Yangian of $\fg^\pm(N)$ with the
deformation parameter equals to $\frac{2\omega}{N-4\theta_0}$.

\section*{Conclusion}

In this letter, we studied new interaction between degree of freedom
for different models with long range interaction such as Calogero
model, Sutherland model or confined Calogero model. For each one, we
obtain the symmetry algebra. Several questions remain open. The Lax
pair as well as the Dunkl operators are not computed for these new
models. These different approaches may allow one to deeply
understand the constraint on the coupling constant which appears
here. Another problem consists in computing the eigenfunctions and
the eigenvalues of these models. The knowledge of the symmetry may
help for their resolution. Indeed, for previous cases like the
Sutherland $gl(N)$-spin model whose the symmetry is the Yangian of
$gl(N)$, the algebra symmetry is crucial to find the spectrum (see
e.g. \cite{BGHP,Hik1,TU}). Finally, we want to point out the problem
of the "freezing" method. Indeed when the coupling constant is not
constrained, it is possible to obtain non-dynamical spin chains
known as Frahm-Polychronakos or Haldane-Shastry spin chains
\cite{Ha,Sh,Po,Fr}. This method seems not to work for the models
studied
in this letter.\\

\textbf{Acknowledgements:} This work is supported by the TMR Network
`EUCLID' Integrable models and applications: from strings to
condensed matter', contract number HPRN-CT-2002-00325.

\appendix

\section{Appendix}

In this appendix, we give some details about the proof of the Serre
relations (\ref{eq:serresut}) and (\ref{eq:serreconfcal}).

\subsubsection*{Sutherland model}

Using relation (\ref{eq:comJ1}), the left-hand side of relation
(\ref{eq:serresut}) can be written as
\begin{equation}
\label{eq:LHSs} LHS={f^{cd,ef}}_{\alpha\beta}
[K_1^{ab},K_1^{\alpha\beta}]+ {f^{ab,cd}}_{\alpha\beta}
[K_1^{ef},K^{\alpha\beta}_1]
+{f^{ef,ab}}_{\alpha\beta}[K_1^{cd},K^{\alpha\beta}_1]\;.
\end{equation}
By direct computation, we obtain
\begin{eqnarray}
\label{eq:k1k1}
[K_1^{ab},K_1^{\alpha\beta}]={f^{ab,\alpha\beta}}_{mn}K_2^{mn}
+X^{ab,\alpha\beta}\;.
\end{eqnarray}
The explicit form of $K_2^{mn}$ is not relevant because the Jacobi
identity for the structure constants implies that these terms vanish
in (\ref{eq:LHSs}). The operator $X^{ab,\alpha\beta}$ takes the
following form,
\begin{eqnarray}
X^{ab,\alpha\beta}&=&-\frac{\lambda}{2}\sum_{j\neq k}
f(x_j,x_k)\left(\theta_a\theta_{\bar \beta}E_j^{\alpha\bar
a}E_k^{{\bar \beta}b}-\theta_\alpha\theta_{\bar b}E_j^{a\bar
\alpha}E_k^{{\bar b}\beta}+\theta_b\theta_{\alpha}E_j^{\bar b\bar
\alpha}E_k^{a\beta}-\theta_a\theta_{\beta}E_j^{\alpha b}E_k^{{\bar
\beta}\bar a}\right)\nonumber\\
&&+\frac{\lambda^2}{4}\sum_{j\neq k \neq \ell}\left(F_j^{\alpha
b}(F_kF_\ell)^{a\beta}-F_j^{a \beta}(F_kF_\ell)^{\alpha
b}+\theta_{\alpha}\theta_{\beta} F_j^{a \bar
\alpha}(F_kF_\ell)^{{\bar \beta} b}-\theta_{\alpha}\theta_{\beta}
F_j^{{\bar \beta} b}(F_kF_\ell)^{a {\bar
\alpha}}\right),\hspace{1cm}
\end{eqnarray}
where
\begin{eqnarray}
f(x_j,x_k)=\left(\frac{\lambda(N-4\theta_0)(x_j+x_k)^2-8
x_jx_k}{2(x_j-x_k)^2}\right)\;.
\end{eqnarray}
Since the right-hand side of (\ref{eq:serresut}) does not depend on
the position, the function $f(x_j,x_k)$ must be constant. This
constraint implies that $\lambda=\frac{2}{N-4\theta_0}$ and
$f(x_j,x_k)=1$.

Now, let us focus on the right-hand side. It contains a sum of
$8^4=4096$ terms. By hand, it would be about impossible to deal with
this number of terms. Fortunately, formal computations with Maple
allow us to reduce this number. Finally, the right-hand side can be
written as
\begin{eqnarray}
\label{eq:RHSs} RHS=\lambda^2\big({f^{cd,ef}}_{\alpha\beta}
Y^{ab,\alpha\beta}+ {f^{ab,cd}}_{\alpha\beta} Y^{ef,\alpha\beta}
+{f^{ef,ab}}_{\alpha\beta} Y^{cd,\alpha\beta}\big)\;,
\end{eqnarray}
where
\begin{equation}
Y^{ab,\alpha\beta}=\sum_{\ell=1}^N\left(\{K_0^{\alpha
b},K_0^{a\ell},K_0^{\ell\beta}\}-\{K_0^{a\beta},K_0^{\alpha
\ell},K_0^{\ell b}\}+\theta_\alpha \theta_\beta\{K_0^{a \bar
\alpha},K_0^{\bar \beta \ell},K_0^{\ell
b}\}-\theta_\alpha\theta_\beta \{K_0^{\bar \beta
b},K_0^{a\ell},K_0^{\ell\bar \alpha}\}\right)\;.
\end{equation}
We recall that $\{.,.,.\}$ is defined by relation (\ref{eq:sym}). To
simplify the notation, we introduce also new generators, for
$(a,b)\in\cE^\pm$,
\begin{equation}
K_0^{\bar b \bar a}=-\theta_a\theta_b K_0^{ab}\;.
\end{equation}
Finally, we compute $Y^{ab,\alpha\beta}$ using explicit expression
(\ref{eq:gen_sut0}) of $K_0^{ab}$ to show that RHS given by
(\ref{eq:RHSs}) is equal to LHS given by (\ref{eq:LHSs}) which
finishes the proof of the Serre relation (\ref{eq:serresut}).

\subsubsection*{Confined Calogero model}

The computation of relation (\ref{eq:serreconfcal}) is simplified by
remarking that its right-hand side is identical to the one of
relation (\ref{eq:serresut}) (up to a factor $4\omega^2$) which has
been computed previously. Its left-hand side (divided by
$-\omega^2$) can be reduced to
\begin{equation}
{f^{cd,ef}}_{\alpha\beta}
\big([J_2^{ab},\cO^{\alpha\beta}_2]+[\cO^{ab}_2,J_2^{\alpha\beta}]\big)
+{f^{ab,cd}}_{\alpha\beta}
\big([J_2^{ef},\cO^{\alpha\beta}_2]+[\cO^{ef}_2,J_2^{\alpha\beta}]\big)
+{f^{ef,ab}}_{\alpha\beta}
\big([J_2^{cd},\cO^{\alpha\beta}_2]+[\cO^{cd}_2,J_2^{\alpha\beta}]\big)
\end{equation}
by remarking, in particular, that the generators $\cO^{ab}_2$ and
$J_2^{ab}$ satisfy the Serre relations (\ref{eq:serrecal}). Now, by
direct computation of the commutator
$[J_2^{ab},\cO^{\alpha\beta}_2]$, using the explicit form
(\ref{eq:J2}) for $J_2^{ab}$, we can prove that the Serre relation
(\ref{eq:serreconfcal}) is satisfied.


\begin{thebibliography}{99}

\bibitem{Cal} F. Calogero, \textsl{Solution of a Three-Body Problem in One
Dimension,} J. Math. Phys. \textbf{10} (1969) 2191. \textsl{Ground
State of a One-Dimensional N-Body System,} J. Math. Phys.
\textbf{10} (1969) 2197. \textsl{Solution of the One-Dimensional
N-Body Problems with Quadratic and/or Inversely Quadratic Pair
Potentials,} J. Math. Phys. \textbf{12} (1971) 419.


\bibitem{Sut} B. Sutherland, \textsl{Exact Results for a Quantum Many-Body
Problem in One Dimension,} Phys. Rev. \textbf{A4} (1971) 2019.
\textsl{Exact Results for a Quantum Many-Body Problem in One
Dimension. II} Phys. Rev. \textbf{A5} (1972) 1372. \textsl{Exact
Ground-State Wave Function for a One-Dimensional Plasma} Phys. Rev.
Lett. \textbf{34} (1975) 1083.

\bibitem{HH} Z.N.C.~Ha and F.D.M.~Haldane,
\textsl{On Models with Inverse-Square Exchange,} Phys. Rev.
\textbf{B46} (1992) 9359 and \texttt{cond-mat/9204017}.

\bibitem{MP} J.A.~Minahan and A.P.~Polychronakos,
\textsl{Integrable Systems for Particles with Internal Degrees of
Freedom,} Phys. Lett. \textbf{B302} (1993) 265 and
\texttt{hep-th/9206046}.

\bibitem{Poly}
A.P. Polychronakos, \textsl{Exchange operator formalism for
integrable systems of particles,} Phys. Rev. Lett \textbf{69} (1992)
703 and \texttt{hep-th/9202057}.

\bibitem{BGHP}
   D. Bernard, M. Gaudin, F.D.M. Haldane, V. Pasquier,
\textsl{Yang-Baxter equation in spin chains with long range
interactions,} J. Phys.  \textbf{A26} (1993) 5219 and
 \texttt{hep-th/9301084}.

\bibitem{BHW} D.~Bernard, K.~Hikami and M.~Wadati,
\textsl{The Yangian Deformation of the W-Algebras and the
Calogero-Sutherland model,} Proceeding of 6-th Nankai Workshop and
\texttt{hep-th/9412194}.

\bibitem{Hik1} K.~Hikami,
\textsl{Symmetry of the Calogero model confined in the harmonic
potential-Yangian and W algebra,} J. Phys. \textbf{A28} (1995) 131,
\textsl{Yangian symmetry and Virasoro character in a lattice spin
with long range interaction,} Nucl. Phys. \textbf{B441} (1995) 530.

\bibitem{F01_1}
F.~Finkel, D.~G{\'o}mez-Ullate, A.~Gonz{\'a}lez-L{\'o}pez,
M.A.~Rodr{\'\i}guez and R.~Zhdanov, \textsl{$A_N$-type Dunkl
operators and new spin Calogero-Sutherland models,} Comm. Math.
Phys. \textbf{221} (2001) 477 and \texttt{hep-th/0102039}.


\bibitem{OP} M.A. Olshanetsky and A.Perelomov,
\textsl{Quantum integrable systems related to the Lie algebras,}
Phys. Rep. \textbf{94} (1983) 313.

\bibitem{caduk}
  V.Caudrelier and N.Cramp\'e,
  \textsl{Integrable N-particle Hamiltonians with Yangian or Reflection Algebra
  Symmetry,}
  J. Phys. \textbf{A37} (2004) 6285 and \texttt{math-ph/0310028}.

\bibitem{ABZABZ78} Al.~B.~Zamolodchikov, Al.~B.~Zamolodchikov,
  \textsl{Relativistic factorized $S$-matrix in two dimensions having $O(N)$
  isotropic symmetry,} Nucl. Phys. \textbf{B133} (1978) 525 and
  \textsl{Factorized $S$-matrices in two dimensions as the exact solutions of
  certain relativistic quantum field models,} Ann.  Phys.  \textbf{120}
  (1979) 253.

\bibitem{KS} P.P.~Kulish, E.K.~Sklyanin,
  \textsl{Solutions of the Yang--Baxter equation,}
  Zap. Nauchn. Sem. LOMI, \textbf{95} (1980) 129 and
  J. Sov. Math. \textbf{19} (1982) 1596.

\bibitem{Isa} A.P. Isaev,
  \textsl{Quantum groups and Yang--Baxter equations,} Phys. Part. Nucl.
  \textbf{26} (1995) 501.

\bibitem{soya}
  D. Arnaudon, J. Avan, N. Cramp\'{e}, L. Frappat and {\'E}. Ragoucy,
  \textsl{$R$-matrix presentation for (super)-Yangians $Y(g)$,}
  J. Math. Phys. \textbf{44} (2003) 302 and \texttt{math.QA/0111325}.

\bibitem{Dri85}
  V.G. Drinfel'd, \textsl{Hopf algebras and the quantum Yang--Baxter
    equation,}
  Soviet.  Math.  Dokl.  \textbf{32} (1985) 254,
  \textsl{Quantum Groups,}
  Proceedings Int. Cong. Math. Berkeley, California, USA (1986) 798,
  \textsl{A new realization of {Y}angians and quantized affine
    algebras,} Soviet.  Math.  Dokl.  \textbf{36} (1988) 212.

\bibitem{TU}
K. Takemura and D. Uglov, \textsl{The orthogonal eigenbasis and
norms of eigenvectors in the Spin Calogero-Sutherland Model,} J.
Phys. \textbf{A30} (1997) 3685 and \texttt{solv-int/9611006}.

\bibitem{Ha} F.D.M. Haldane, \textsl{Exact Jastrow-Gutzwiller
resonating-valence-bond ground state of the spin-1/2
antiferromagnetic Heisenberg chain with 1/r2 exchange,} Phys. Rev.
Lett. \textbf{60} (1988) 635.

\bibitem{Sh} B.S. Shastry, \textsl{Exact solution of an
S=1/2 Heisenberg antiferromagnetic chain with long-ranged
interactions,} Phys. Rev. Lett. \textbf{60} (1988) 639.

\bibitem{Po} A.P. Polychronakos, \textsl{Lattice integrable systems of
 Haldane-Shastry type,} Phys. Rev. Lett. \textbf{70} (1993) 2329 and
 \texttt{hep-th/9210109}.

\bibitem{Fr} H. Frahm, \textsl{Spectrum of a spin chain with inverse square
exchange,} J. Phys \textbf{A26} (1993) L473 and
\texttt{cond-mat/9303050}.

\end{thebibliography}
\end{document}